\documentclass[10pt]{article}

\usepackage[compress]{cite}
\usepackage{amsmath,bm}
\usepackage{hyperref}
\usepackage{enumerate}
\usepackage{bm}
\usepackage{hyperref}
\usepackage{graphicx}
\usepackage[dvipsnames]{xcolor}
\usepackage{amsfonts}
\usepackage{amssymb,amsthm}
\usepackage{mathrsfs}
\usepackage{bbold}
\usepackage{orcidlink}

\usepackage[left=2.5cm,right=2.2cm,top=0.5cm,bottom=1.3cm,includeheadfoot]{geometry}

\newcommand{\ham}{\mathcal{H}}
\newcommand{\diff}{\mathcal{D}}
\newcommand{\shift}{{N^x}}
\newcommand{\lapse}{N}
\newcommand{\erad}{E^x{}}
\newcommand{\ephi}{E^\varphi{}}
\newcommand{\krad}{K_x}
\newcommand{\kang}{K_\varphi}

\newcommand{\qbat}{E^x{}}
\newcommand{\qbi}{E^\varphi{}}

\newcommand{\pbat}{K_x}
\newcommand{\pbi}{K_\varphi}

\newcommand{\mass}{M}
\newcommand{\hs}{U}
\newcommand{\s}{{\sigma}}

\begin{document}
	\begin{center}
		{\Large\bf 
	  Spacetime geometry from canonical spherical gravity\\
		}
		\vskip 5mm
		{\large
			Asier Alonso-Bardaji\orcidlink{0000-0002-8982-0237}$^1$ and David Brizuela\orcidlink{0000-0002-8009-5518}$^2$}
		\vskip 3mm
		{\sl $^1$Aix Marseille Univ., Univ. de Toulon, CNRS, CPT, UMR 7332, 13288 Marseille, France}\\
		\vskip 1mm
		{\sl $^2$Department of Physics and EHU Quantum Center, University of the Basque Country UPV/EHU,\\
			Barrio Sarriena s/n, 48940 Leioa, Spain}
	\end{center}
 	\vskip 2mm
 		\begin{abstract}
 		We study covariant models for vacuum spherical gravity within a canonical setting.
 		Starting from a general ansatz,
 		we derive the most general family of Hamiltonian constraints that are quadratic in
 		first-order and linear in second-order spatial derivatives of the triad variables, and obey certain
 		specific covariance conditions. These conditions ensure that the dynamics
 		generated by such family univocally defines a spacetime geometry, independently
 		of gauge or coordinates choices.
 		This analysis generalizes the Hamiltonian constraint of general relativity, though keeping
 		intact the covariance of the theory, and leads to a rich variety of new geometries.
 		We find that the resulting geometries
 		depend on seven free functions of one scalar variable, and we study their generic features.
 		By construction, there are no propagating degrees of freedom in the theory. However,
 		we also show that it is possible to add matter to the system by simply following
 		the usual minimal-coupling prescription, which leads to novel models to describe
 		dynamical scenarios.
\end{abstract}

\section{Introduction}

The construction of effective models for loop quantum gravity has become a hot topic in the last years, particularly due to the success of such models in cosmological scenarios, where they show a generic resolution of the big bang singularity \cite{Ashtekar:2007em,Agullo:2016tjh,Ashtekar:2006wn,Ashtekar:2011ni,Agullo:2013dla}. More recently, the attention has turned to spherically symmetric scenarios \cite{Modesto:2008im,Boehmer:2007ket,Joe:2014tca,Olmedo:2017lvt,BenAchour:2018khr},
to check whether those predictions of singularity resolution are robust or rather an artifact of excessive symmetry assumptions. These effective models are constructed introducing certain
(so-called holonomy and inverse-triads) modifications
on the Hamiltonian constraint of general relativity (GR). However, it is widely known that the covariance of the theory is not explicit in the canonical setting \cite{Teitelboim:1972vw,Pons:1996av}.
Therefore, in general,
such modifications will
make the predictions of the model to depend on the specific gauge or coordinate
choice.
This is certainly the case in \cite{Kelly:2020uwj,Ashtekar:2018cay,Ashtekar:2018lag,Bodendorfer:2019cyv,Bodendorfer:2019nvy,Ashtekar:2020ckv,Bojowald:2018xxu,Gambini:2020nsf}, where the effective geometries are defined with a particular (gauge-fixed) solution of the Hamilton equations,
and a change of gauge leads, in general, to a different geometry.

Attempts to construct covariant models in this context led to no-go results, suggesting
that holonomy corrections were not compatible with covariance, particularly when coupling matter
\cite{Tibrewala:2013kba,Bojowald:2015zha,Bojowald:2020dkb,Bojowald:2020unm
,Alonso-Bardaji:2020rxb}.
Nevertheless, these conclusions were ruled out by the explicitly covariant models studied in
detail in Refs.~\cite{Alonso-Bardaji:2021yls,Alonso-Bardaji:2022ear,Alonso-Bardaji:2023niu}.
In fact, these are the only effective covariant (in the sense that any gauge choice provides the
same geometry) models in the literature to describe
a spherical black hole with corrections motivated by loop quantum gravity, both in vacuum
\cite{Alonso-Bardaji:2021yls,Alonso-Bardaji:2022ear} and including
charge and a cosmological constant \cite{Alonso-Bardaji:2023niu}.

The aim of the present work is to perform a systematic study of the covariance of all possible
modifications to the vacuum GR Hamiltonian constraint that do not increase the derivative order
of the theory. For such a purpose, we will perform an analysis similar to the one presented
in \cite{Alonso-Bardaji:2021tvy} for spherical models coupled to matter. There, starting from a generic ansatz,
we derived a modified Hamiltonian constraint that forms a closed hypersurface deformation algebra
with the diffeomorphism constraint. However, in order to provide a geometry in an unambiguous way,
this is not enough, and one needs to ensure that the structure functions of such algebra
have the correct transformation properties (this was indeed the case in the vacuum models studied
in \cite{Alonso-Bardaji:2021yls,Alonso-Bardaji:2022ear,Alonso-Bardaji:2023niu},
but not in general for the models coupled to matter presented in \cite{Alonso-Bardaji:2021tvy}).

Therefore, we will propose an ansatz for the modified vacuum Hamiltonian constraint,
which will be quadratic in first-order and linear in second-order spatial derivatives of the triad variables,
while allowing a generic dependence on all the variables of the model.
Then, two conditions will be implemented, the closure of the canonical hypersurface deformation
algebra and the embeddability condition \cite{Teitelboim:1972vw}, which will ensure the covariance of the model.
These conditions will greatly restrict the form of the possible modifications, and we will obtain
the most general Hamiltonian constraint that obeys such conditions. Then, we will explicitly
provide the form of the metric described by such constraints, and will analyze its main
geometric features. We would like to point out that the main results of this study
{were already available in \cite{AlonsoBardaji:2023bww}}.

The article is organized as follows. In Sec. \ref{sec:gr} we review some basic notions of spherical GR in its canonical formulation. In Sec.~\ref{sec.conditions} we set our requirements for the modified models that we will implement in Sec.~\ref{sec.constr}. The reader interested in the final result of the computation may jump directly from Sec.~\ref{sec.conditions} to Sec.~\ref{sec:result}, where the main
result of the analysis is presented. In Sec.~\ref{sec.st} we study some generic features of all the
spacetime geometries described in the article. We also show how to minimally couple matter fields to the modified vacuum models in Sec.~\ref{sec.matter}, and we end the manuscript by summarizing and discussing our main results in Sec.~\ref{sec.concl}.

\section{Spherical vacuum in canonical general relativity}\label{sec:gr}

In spherical symmetry, the spacetime manifold ${\cal M}$ is a warped product between a two-dimensional
manifold ${\cal M}^2$ and the two-sphere $S^2$. Choosing coordinates adapted to the symmetry, the spacetime metric is diagonal by blocks:
\begin{align}\label{eq.metricgen}
    ds^2=g_{AB}dy^A dy^B+r^2 d\Omega^2,
\end{align}    
where $d\Omega^2$ is the metric of the two-sphere, and capital Latin indices take values $0$ and $1$.
(When convenient, we will denote the coordinates as $t:=y^0$ and $x:=y^1$.)
The area-radius function $r=r(t,x)$ is a scalar field on the manifold ${\cal M}^2$
and encodes the area of the spheres of constant $r$.

Due to the symmetry, the angular components of the diffeomorphism constraint {of general relativity}
are vanishing, while the radial component reads,
\begin{equation}\label{eq.diff}
 \diff=-E^{x\prime}K_x+E^\varphi K_\varphi',
\end{equation}
where the prime stands for a derivative with respect to $x$.
{
These are canonically conjugate variables, which, on a spatial leaf with a constant value of $t$, obey}
\begin{align}
\{K_x(t,x_1), E^x(t,x_2)\}=\{K_\varphi(t,x_1),E^\varphi(t,x_2)\}=\delta(x_1-x_2).
\end{align}
In terms of these variables, the Hamiltonian constraint of general relativity
{(assuming a vanishing cosmological constant)}
reads,
\begin{align}\label{eq.hamGR}
 \ham^{\rm (GR)}=& -\frac{{{E}^\varphi}}{2\sqrt{{{E}^x}}}\left(1+{K}_\varphi^2\right)  -2\sqrt{{{E}^x}}{{K}_x}{{K}_\varphi}
+\frac{({E}^x{}')^2}{8\sqrt{{E}^x}{E}^\varphi}
            -\sqrt{{E}^x}\frac{{E}^x{}'{E}^\varphi{}'}{2({E}^\varphi)^2}  +\sqrt{{E}^x}\frac{{E}^x{}''}{2{E}^\varphi}.
\end{align}
As it is well known, these constraints obey the
hypersurface deformation algebra,
\begin{subequations}\label{eq.hdageneral0}
  \begin{align}
  \label{eq.ddqp0}  \big\{D[s_1],D[s_2]\big\}&=D\big[s_1s_2'-s_1's_2\big],\\
 \label{eq.dhqp0} \big\{D[s_1],{H}^{\rm (GR)}[s_2]\big\}&={H}^{\rm (GR)}\big[s_1s_2'\big],\\
  \label{eq.hhqp0}   \big\{{H}^{\rm (GR)}[s_1],{H}^{\rm (GR)}[s_2]\big\}&=  D\big[
    \,q^{xx}(s_1s_2'-s_1's_2)\big],
  \end{align}
\end{subequations}
where $H^{\rm (GR)}[s]:=\int s\ham^{\rm (GR)} dx $ and $D[s]:=\int s\diff dx$ are the smeared forms of the constraints, and $q^{xx}:=E^x/(E^\varphi)^2$ {is positive, which signals that
the spacetime is Lorentzian}.

In this canonical setting,
the algebra encodes the covariance of the theory. Since they commute on-shell, $\ham$ and $\diff$ are first-class constraints,
and thus generators of gauge transformations on the phase space.
These gauge transformations also describe coordinate
transformations on the spacetime manifold, and one can indeed understand
the action of $\diff$ as generating deformations on the spatial leaf,
while $\ham$ generates deformations along the normal direction. 
Besides, from the structure function in \eqref{eq.hhqp0},
one can reconstruct the metric on $\mathcal{M}^2$,
\begin{align}\label{eq.metric1}
    g_{AB}dy^A dy^B={\alpha}\lapse^2 dt^2+\frac{1}{|q^{xx}|}{\big({dx}+N^x{dt}\big)^2},
\end{align}
where $N$ and $N^x$ are the Lagrange multipliers that define the total
Hamiltonian $H_T=H^{\rm (GR)}[N]+D[N^x]$, which generates the dynamics,
and $\alpha:=-\mathrm{sgn}(q^{xx})$ is the signature of the spacetime manifold \cite{Teitelboim:1972vw}.
{
We have included the absolute value $|q^{xx}|$ in the metric,
though in the GR case it is not needed since $q^{xx}$ is positive, $\alpha=-1$, and thus
the manifold is Lorentzian. However, in this form, expression \eqref{eq.metric1}
would also be valid for a negative $q^{xx}$, with $\alpha=1$, which would lead to a Riemannian metric. 
}

At this point we note that the algebra \eqref{eq.hdageneral0} does not have
any information about the angular components of the metric, and, in particular,
about the scalar $r$. This is due to the symmetry reduction. {For}
the full (nonreduced) theory, one can {indeed} read out the inverse of the whole spatial
metric from the bracket \eqref{eq.hhqp0}.

\section{Covariant deformations of the Hamiltonian constraint}\label{sec.conditions}

In brief,
the goal of this paper is to construct the most general Hamiltonian constraint
$\ham$ with the same derivative structure as the GR constraint \eqref{eq.hamGR},
with a well defined limit to GR,
and such that
it covariantly defines a spacetime metric {along with the diffeomorphism constraint \eqref{eq.diff}}. These requirements
can be translated into four precise conditions.

\textbf{\textit{(i) Derivative structure.}}
The first condition concerns the derivative structure:
we will not consider high-order derivative corrections to the GR constraint \eqref{eq.hamGR}. 
Therefore, the most general constraint that is quadratic in derivatives of
$(E^x, E^\varphi)$, and linear in their second derivatives, reads
\begin{align}
   \label{eq.hamqpvac} \ham=&\,a_{0} +a_1\erad''+a_2\ephi''
   + a_{11}(\erad')^2+ a_{12}\erad'\ephi'+ a_{22}(\ephi')^2,
\end{align}
where the six functions $a_0$, $a_1$, $a_2$, $a_{11}$, $a_{12}$, and $a_{22}$
are completely free functions of the four variables
$(\erad, \ephi,\krad, \kang)$, but they do not depend on their derivatives.

\textbf{\textit{(ii) GR limit.}} The second condition {will ensure} that
GR is recovered as a continuous limit of the model. {For such a purpose}, we will demand
that the functions $a_0$, $a_1$, $a_{11}$, and $a_{12}$ are not identically vanishing, so that
\eqref{eq.hamGR} is {automatically} included as a particular case of
\eqref{eq.hamqpvac}.

\textbf{\textit{(iii) {Closure of the canonical} hypersurface deformation algebra.}}
{
{For t}he theory to be covariant, the hypersurface deformation algebra should be closed and, thus, there should
be no anomalies (terms on the right-hand side of the brackets between the constraints
\eqref{eq.diff} and \eqref{eq.hamqpvac} that do not vanish on the constraint surface ${\cal H}=0={\cal D}$).
In addition,}
in order to interpret the constraint \eqref{eq.hamqpvac} as the generator of deformations
along the normal direction to the spatial leaves, it needs
to obey the canonical form of the algebra,
\begin{subequations}\label{eq.hdageneral}
  \begin{align}
  \label{eq.ddqp}  \big\{D[s_1],D[s_2]\big\}&=D\big[s_1s_2'-s_1's_2\big],\\
 \label{eq.dhqp} \big\{D[s_1],{H}[s_2]\big\}&={H}\big[s_1s_2'\big],\\
  \label{eq.hhqp}   \big\{{H}[s_1],{H}[s_2]\big\}&=D\left[
    F(s_1s_2'-s_1's_2)\right],
  \end{align}
\end{subequations}
with ${H}[{s}]:=\int {s}{\cal H}dx$ and a certain {nonexactly vanishing} structure function $F$.

\textbf{\textit{(iv) Spacetime embeddability.}}
Given an algebra of the form \eqref{eq.hdageneral}, {and} following the discussion
of the previous section, one would be tempted to write the associated metric as
\begin{align}\label{eq.metric}
    g_{AB}dy^A dy^B=\sigma\lapse^2 dt^2+\frac{1}{|F|}{\big({dx}+N^x{dt}\big)^2},
\end{align}
where $\sigma:=-{\rm{sgn}}(F)$, while $\lapse$ and $\shift$ are the Lagrange multipliers that define the total
Hamiltonian $H_T=H[\lapse]+D[\shift]$.
However, this will covariantly define a metric tensor on spacetime
only if the structure function $F$
has the adequate transformation properties.
Let us see this in more detail.

Gauge transformations on the phase space are generated by {the generator} $H[\epsilon^\bot]+D[\epsilon^\parallel]$,
for certain gauge parameters $\epsilon^\bot$ and $\epsilon^\parallel$. Since $H$ generates
normal and $D$ tangential deformations with respect to the spatial leaf,
such a gauge transformation must be equivalent to the infinitesimal
coordinate transformation {implemented by
${\cal L}_\epsilon g_{AB}$,where
the vector $\epsilon$ is given in terms of the gauge parameters as
$\epsilon^A\partial_A=
\epsilon^\bot n^A\partial_A +\epsilon^\parallel\partial_x$
}, with $n^A\partial_A:=(\partial_t-N^x\partial_x)/N$ being the unit normal to the spatial leaf.

Therefore, the last condition we will require is that the structure function $F$ transforms
adequately, in order to {consistently} interpret it as the inverse of the spatial metric. This condition is expressed in a simpler way if we use the $(t,x)$ components of the vector $\epsilon$ rather than the normal-tangential ones, i.e., $\epsilon^A\partial_A=\epsilon^t\partial_t+\epsilon^x\partial_x$, where the relations $\epsilon^\bot=\epsilon^t N$ and $\epsilon^\parallel=\epsilon^tN^x+\epsilon^x$ can be directly obtained using the definition of the unit normal. Then,
the embeddability condition {reads}
\begin{align}\label{eq.covarianceoneoverf}
&\epsilon^t\partial_t\left({{1}/{F}}\right)+\epsilon^x\partial_x \left({1}/{F}\right) +\left(2/F\right)\left({N^x}\partial_x{\epsilon^t}+\partial_x{\epsilon^x}\right)
\approx\left\{\left({1}/{F}\right),H\big[ \epsilon^t N\big]+D\big[\epsilon^tN^x+\epsilon^x\big]\right\},
\end{align}
{where the left-hand side is simply the transformation of $F$ as the inverse of the spatial
metric under the Lie dragging ${\cal L}_\epsilon g_{AB}$, while the right-hand side provides
the gauge transformation of $F$ as a phase-space function with the corresponding gauge parameters
(for more explicit details about this derivation see, e.g., Sec. 3 of Ref. \cite{Alonso-Bardaji:2022ear}).
Note that this condition should} be satisfied for any $\epsilon^t$ and $\epsilon^x$ on-shell, i.e.,
when the constraints vanish ${\cal H}=0={\cal D}$, and the equations of motion
\begin{align}
\dot{E}^x=\{E^x,H[N]+D[N^x]\},\\
\dot{K}_x=\{K_x,H[N]+D[N^x]\},\\
\dot{E}^\varphi=\{E^\varphi,H[N]+D[N^x]\},\\
\dot{K}_\varphi=\{K_\varphi,H[N]+D[N^x]\},
\end{align}
are obeyed. {Here,} the dot stands for the derivative with respect to $t$.
All along the paper we will use the symbol $\approx$ for an equality that is obeyed on-shell,
though in some cases it will also mean on the constraint surface (${\cal H}=0={\cal D}$).

In summary, we will require that the Hamiltonian constraint \eqref{eq.hamqpvac}, along with the diffeomorphism
constraint \eqref{eq.diff}, should obey the canonical form
of the algebra \eqref{eq.hdageneral} and the embeddability condition \eqref{eq.covarianceoneoverf},
with the functions $a_0$, $a_1$, $a_{11}$, and $a_{12}$
being not exactly vanishing.

\section{Construction}\label{sec.constr}

In this section we will construct the most general Hamiltonian constraint that obeys
the four conditions {\it (i)-(iv)} detailed in the previous section. More precisely,
beginning from the ansatz \eqref{eq.hamqpvac}, in Sec. \ref{sec:anomaly_freedom} we will
impose the requirement {of the closure of the canonical algebra by ensuring that \eqref{eq.hdageneral}
is obeyed}. This will significantly reduce the freedom of the free functions, by completely fixing the functional dependence of the Hamiltonian constraint on the variables
$\krad$ and $\ephi$. Once this is done, in Sec. \ref{sec:embeddability}, we will impose that the structure function
$F$ obeys the spacetime embeddability condition \eqref{eq.covarianceoneoverf}. As it will be shown below, this requirement will only leave certain freedom on the functional dependence on $\erad$.
Along the way, following condition {\it (ii)}, we will disregard any solution that
imposes an exactly vanishing value of any of the functions $a_0$, $a_1$, $a_{11}$, or $a_{12}$.

The reader interested only in the final result can skip the following two subsections and go directly
to Sec. \ref{sec:result}, where Eq. \eqref{hamSO3vacmod+} displays
the most general Hamiltonian constraint that obeys conditions {\it (i)-(iv)}.

\subsection{{Closure of the canonical hypersurface deformation algebra}}\label{sec:anomaly_freedom}

Since we are considering the classical form of the diffeomorphism constraint \eqref{eq.diff},
one can easily check that the bracket of the diffeomorphism constraint with itself follows \eqref{eq.ddqp}, and no anomalies arise. Let us now compute the nontrivial brackets
$\{D[s_1],H[s_2]\}$ and $\{H[s_1],H[s_2]\}$. 
We note that, in order to find the canonical form of the algebra
\eqref{eq.hdageneral}, in several instances we will need to integrate by parts.
In such cases we will simply drop out the total derivatives from the integrand,
and thus neglect the boundary terms.

\subsubsection{The bracket $\{D[s_1],H[s_2]\}$}

One can directly take the form of the Hamiltonian constraint \eqref{eq.hamqpvac}, compute $\{D[s_1],H[s_2]\}$,
and find restrictions on the free functions by imposing that the bracket is given
by \eqref{eq.dhqp}. However, the computations can be somehow simplified a bit, if one first considers
the geometric meaning of this bracket.

The diffeomorphism constraint is the generator of gauge transformations on the spatial
leaf. More precisely, a gauge transformation generated by $D[u^x]$ on a phase-space
function $f$ corresponds
to a Lie dragging of $f$ along the vector $u=u^x\partial_x$ on spacetime, that is,
$\{f,D[u^x]\}={\cal L}_u f$.
Therefore, the bracket $\{D[s_1],H[s_2]\}$ can be understood as the gauge transformation of
the Hamiltonian constraint and, if it is given by \eqref{eq.dhqp}, it simply
means that the Hamiltonian constraint is a weight-one scalar density\footnote{Recall that the Lie derivative of a weight-$w$ scalar density $f$
is given by ${\cal L}_u f= u^x f'+ w f {u^{x}}'$.}.
Let us thus analyze the weight of the different variables.
The gauge transformations,
{
\begin{align}
\{\pbat,D[u^x]\} &=u^x\pbat'+{u^x}'\pbat,\\
\{\qbat,D[u^x]\} &=u^x\qbat',\\
\{\pbi,D[u^x]\} &=u^x\pbi',\\
\{\qbi,D[u^x]\} &=u^x\qbi'+{u^x}'\qbi,
\end{align}
}%
imply that $E^x$ and $K_\varphi$ are (weight-0) scalars
on the spatial leaf,
while $E^\varphi$ and $K_x$ are weight-one scalar densities.
Therefore, without considering derivatives,
there are three independent scalar quantities, and the most general scalar function on the spatial
leaf is given by $f(\pbat/\qbi,\qbat,\pbi)$.

Now, the action of the derivative with respect to $x$ on a scalar density
will increase its weight by one, though it generically will produce an object
that is not a scalar density.
This is easy to
see by taking into account that, in any one-dimensional manifold (like the
spatial leaf under consideration),
a weight-$w$ scalar density is equivalent to a (weight-0)
covariant tensor of rank $w$\footnote{Here we are assuming a positive $w$. A scalar density with a negative weight $w$
is equivalent to a (weight-0) contravariant tensor of rank $|w|$.}.
Therefore, the derivative of a scalar, like for instance
$E^{x}{}'$, is automatically a weight-one scalar density (i.e., a one-form).
However, the derivative of a weight-one scalar density (i.e., a one-form), like
for instance $E^{x}{}''$ or $E^{\varphi}{}'$, will not have
the correct transformation properties of
a weight-two scalar density (i.e., a rank-two covariant tensor). 

With this information at hand, we construct,
among the family of Hamiltonian constraints \eqref{eq.dhqp}, the most general
weight-one expression {(i.e., counting every prime as contributing one to the weight)} as
\begin{align}
\label{eq.hamqpb} 
\ham&= \qbi b_{0} 
+b_1\frac{\erad''}{\ephi}+b_2\frac{\ephi''}{\ephi^2}
   + b_{11}\frac{(\erad')^2}{\ephi}+b_{12}\frac{\erad'\ephi'}{\ephi^2}+ b_{22}\frac{(\ephi')^2}{\ephi^3},
\end{align}
with $b_k=b_k(\pbat/\qbi,\qbat,\pbi)${, for $k=0,1,2,11,12,22$,} being free functions of the three independent
scalar quantities (on the spatial leaf) of the model. However, as expected, \eqref{eq.hamqpb}
is not a weight-one scalar density (i.e., a one-form),
and there are still certain restrictions that must be imposed so that
\eqref{eq.dhqp} is obeyed.

More precisely, if one computes
the Poisson bracket between the diffeomorphism constraint \eqref{eq.diff} and
the Hamiltonian constraint \eqref{eq.hamqpb},
after removing derivatives of $s_1$ through integration by parts,
one can write
\begin{align}
    \big\lbrace D[s_1],{H}[s_2]\big\rbrace = \int{dx} \,s_1\left(s_2\Gamma_0 +s_2'\Gamma_1 +s_2''\Gamma_2\right).
\end{align}
Here $\Gamma_0$, $\Gamma_1$, and $\Gamma_2$ are complicated
expressions that depend on the six free functions $b_k$ and their partial derivatives,
and also on all the variables $(E^x,K_x,E^\varphi,K_\varphi)$
and their radial derivatives.
However, the terms $\Gamma_0$, $\Gamma_1$, and $\Gamma_2$ do not depend on $s_2$ nor its derivatives.
Therefore, the right-hand side of the above expression will vanish on the constraint surface,
$s_2\Gamma_0 +s_2'\Gamma_1 +s_2''\Gamma_2\approx0$, only if each of the terms
vanish independently, that is,
\begin{align}
 \Gamma_0 &\approx 0,\\
  \Gamma_1 &\approx 0,\\
   \Gamma_2 &\approx 0.
\end{align}
In order to evaluate the different expressions on the constraint surface, we isolate $\qbat''$ and $\pbi'$ from the constraint
equations ${\cal D}\approx 0$ and ${\cal H}\approx 0$, with their form given in
\eqref{eq.diff} and \eqref{eq.hamqpb}, respectively. Then, we substitute $\erad''$, $\kang'$, and their subsequent derivatives in $\Gamma_0$, $\Gamma_1$, and $\Gamma_2$. Last, we remove all the terms including $\diff$, $\ham$, and their derivatives. This procedure leads to
\begin{align}\label{eq.gamma2}
  \Gamma_2\approx   {\mathcal{A}}^{(2)}_{1}\pbat' +{\mathcal{A}}^{(2)}_{2}\qbat'+{\mathcal{A}}^{(2)}_{3}\pbi',
\end{align}
with
\begin{align}
    {\mathcal{A}}^{(2)}_{1}&:=\frac{3}{\qbi^2}\frac{\partial b_{2}}{\partial\pbat},\\
    {\mathcal{A}}^{(2)}_{2}
    &:=\frac{1}{\qbi}\left(3\frac{\pbat}{\qbi}\frac{\partial b_{2}}{\partial\pbi} +3\frac{\partial b_{2}}{\partial\qbat}-b_{12}-b_{1} \right),\\
    {\mathcal{A}}^{(2)}_{3}&:=-\frac{1}{\qbi^2}\left(2 b_{22}+6b_{2}+3\frac{\pbat}{\qbi} \frac{\partial b_{2}}{\partial \pbat}\right).
\end{align}
Now, since the free functions $b_k$ do not depend on the derivatives {of the variables with respect to $x$},
$\Gamma_2\approx 0$ will be obeyed only if
the coefficient of any primed variable
in {expression \eqref{eq.gamma2}} vanishes
by itself \textit{off-shell},
that is, 
\begin{align}
    \Gamma_2\approx0\quad\iff\quad{\mathcal{A}}^{(2)}_{i}=0\qquad\forall\quad i=1,2,3 .
\end{align}
We have thus found three independent equations, leading to the following three conditions,
\begin{align}
    \frac{\partial b_2}{\partial\pbat}&=0,\label{condition1}\\b_{22} &=-3b_{2}\label{condition2},\\
        b_{12}&=-b_1+3\frac{\pbat}{\qbi}\frac{\partial b_{2}}{\partial\pbi} +3\frac{\partial b_{2}}{\partial\qbat}.\label{condition3}
\end{align}
Enforcing these conditions in $\Gamma_1$, its form is simplified a bit,
and {it reads}
\begin{align}
    \Gamma_1\approx&\;
  {\mathcal{A}}^{(1)}_{1} +{\mathcal{A}}^{(1)}_{2}\pbat'\qbat'+{\mathcal{A}}^{(1)}_{3}(\qbat')^2
 +{\mathcal{A}}^{(1)}_{4}\qbat'\qbi'+{\mathcal{A}}^{(1)}_{5}(\qbi')^2+{\mathcal{A}}^{(1)}_{6}\qbi'',
\end{align}
{
where all the derivatives of the variables are written explicitly. Therefore, using the same rationale as above,}
\begin{align}
    \Gamma_1\approx0\quad\iff\quad{\mathcal{A}}^{(1)}_{i}=0\qquad\forall\quad i=1,\dots,6 .
\end{align}
We proceed to read the easiest conditions and use them to simplify the remaining anomalies. First,
\begin{align}
    0={\mathcal{A}}^{(1)}_{2}=-\frac{3}{\qbi^2}\frac{\partial b_2}{\partial\pbi}
\end{align}
demands that $b_2$ is independent of $\pbi$. Using this,
\begin{align}
    0={\mathcal{A}}^{(1)}_{1}=3\qbi \frac{b_0}{b_1}\frac{\partial b_2}{\partial\qbat}
\end{align}
requires that $b_2$ is a constant function. Finally,
\begin{align}
    0={\mathcal{A}}^{(1)}_{6}= \frac{3b_2}{\qbi^2}
\end{align}
sets $b_2=0$, thus removing the term $\ephi''$ from the Hamiltonian.
In fact, one can check that these three conditions, along with \eqref{condition1}--\eqref{condition3},
are necessary and sufficient so that, not only $\Gamma_1\approx0$
is satisfied, but also $\Gamma_0\approx0$.

In order to display the final form of the constraint,
for convenience, let us define $\tilde{b}_0:=b_0/b_1$ and $\tilde{b}_{11}:=b_{11}/b_1$ (recall that, due to
condition {\it (ii)}, $b_1$ cannot be identically vanishing). In this way,
after implementing the above conditions on \eqref{eq.hamqpb},
the resulting constraint is given by
\begin{align}\label{eq.hamgdh}
    \ham&=b_1\left(\qbi \tilde{b}_{0} +\tilde{b}_{11}\frac{({\qbat'})^2}{\qbi}-\frac{{\qbat'}{\qbi'}}{\qbi^2}+\frac{\qbat''}{\qbi}\right),
\end{align} 
which yields \eqref{eq.dhqp} along with \eqref{eq.diff}, that is, $\Gamma_0=0$, $\Gamma_1=1$, and $\Gamma_2=0$ \textit{off-shell}.
Note that,
at this point of the analysis, there are only three free functions
$\tilde{b}_0$, $\tilde{b}_{11}$, and $b_{1}$,
which depend on the three scalar combinations of the variables
$\pbat/\qbi$, $\pbi$, and $\qbat$.

\subsubsection{The bracket $\{H[s_1],H[s_2]\}$}

We now move on to compute the Poisson bracket between
two Hamiltonian constraints \eqref{eq.hamgdh}, and impose the
{necessary conditions so that its form is given by \eqref{eq.hhqp}}.
If we define $s:=s_1s_2'-s_1's_2$,
and remove all derivatives of $s$ through integration by parts, the bracket takes the form
\begin{align}\label{hhbracketfv}
    \big\lbrace H[s_1],{H}[s_2]\big\rbrace = \int{dx} \,(s_1s_2'-s_1's_2)\Gamma_{3},
\end{align}
where $\Gamma_3$ does not depend on $s_1$ nor $s_2$,
and it is a combination of the different variables
$(E^x,K_x,E^\varphi,K_\varphi)$ and their first-order {derivatives $({E^x}',{K_x}',{E^\varphi}',{K_\varphi}')$}.
{As above,}
we use now \eqref{eq.diff} and \eqref{eq.hamgdh} to solve $\diff\approx0$ and $\ham\approx0$ for $\pbi'$ and $\qbat''$, respectively. Substituting them in $\Gamma_3$ and setting the constraints to zero, we find
\begin{align}
    \Gamma_3\approx&\;
 {\mathcal{A}}^{(3)}_{1}\pbat'+{\mathcal{A}}^{(3)}_{2}\qbat'+{\mathcal{A}}^{(3)}_{3}\qbi'+{\mathcal{A}}^{(3)}_{4}\pbat'(\qbat')^2
+{\mathcal{A}}^{(3)}_{5}\qbi'(\qbat')^2+{\mathcal{A}}^{(3)}_{6}(\qbat')^3.
\end{align}
In this expression all derivatives of the variables with respect to $x$ are explicit, and thus the 
anomalies ${\mathcal{A}}\,{}^{(3)}_{i}$ depend on the variables $(E^x,K_x,E^\varphi,K_\varphi)$,
but not on their derivatives. Therefore, as done in the previous subsection,
the condition of anomaly freedom is translated to the vanishing of every  ${\mathcal{A}}\,{}^{(3)}_{i}$ coefficient, that is,
\begin{align}
    \Gamma_3\approx0\quad\iff\quad{\mathcal{A}}^{(3)}_{i}=0\qquad\forall\quad i=1,\dots,6 .
\end{align}
The simplest conditions are given by 
\begin{subequations}
    \begin{align}
        0={\mathcal{A}}^{(3)}_{1}=-\frac{\qbi}{\pbat}{\mathcal{A}}^{(3)}_{3}& =-b_1^2\frac{\partial^2 \tilde{b}_0}{\partial \pbat^2},\\[6pt]
        0={\mathcal{A}}^{(3)}_{4}=-\frac{\qbi}{\pbat}{\mathcal{A}}^{(3)}_{5}&=-\frac{b_1^2}{\qbi^2}\frac{\partial^2 \tilde{b}_{11}}{\partial \pbat^2} ,    
    \end{align}
\end{subequations}
implying that both $\tilde{b}_0$ and $\tilde{b}_{11}$ are at most linear in $\pbat$,
\begin{subequations}
    \begin{align}
    \tilde{b}_0&=c_{00}(\pbi,\qbat)+\frac{\pbat}{\qbi} c_{01}(\pbi,\qbat),\\[4pt]
    \tilde{b}_{11}&=c_{10}(\pbi,\qbat)+\frac{\pbat}{\qbi} c_{11}(\pbi,\qbat).
\end{align}
\end{subequations}
After enforcing these conditions, the remaining two anomalies are simplified to
\begin{subequations}
\begin{align}
     0={\mathcal{A}}^{(3)}_{6}&=\frac{b_1^2}{\qbi^3}\left(\frac{\partial c_{10}}{\partial \pbi}-\frac{\partial c_{11}}{\partial \qbat}\right),\\[6pt]
        \!\!0={\mathcal{A}}^{(3)}_{2}&=\frac{b_1^2}{\qbi}\!\left(\!\frac{\partial c_{00}}{\partial {\pbi}} +\! 2(c_{00}c_{11}-c_{01}c_{10})-\!\frac{\partial c_{01}}{\partial {\qbat}}\!\right)\!.\!   
\end{align}
\end{subequations}
It is easy to see that
the general solution to these equations
can be written in terms of two free functions of two variables, $f(\pbi,\qbat)$ and $g(\pbi,\qbat)\neq0$, and two additional functions, $u(\qbat)$ and $v(\qbat)$, which only depend on $\qbat$, as follows,
\begin{subequations}
    \begin{align}
    c_{00}&=\frac{1}{g}\bigg(\frac{\partial f}{\partial\qbat}+v +u f \bigg),\\
    c_{01}&=\frac{1}{g}\frac{\partial f}{\partial\pbi},\\[6pt]
    c_{10}&=\frac{u}{2}+\frac{1}{2}\frac{\partial}{\partial\qbat}\Big[\log\big(g\big)\Big],\\[6pt]
    c_{11}&=\frac{1}{2}\frac{\partial}{\partial\pbi}\Big[\log\big(g\big)\Big].
\end{align}
\end{subequations}
Therefore, replacing the above conditions in \eqref{eq.hamgdh}, we obtain
\begin{align}\label{eq.hamU}
    &\ham={b}_1\Bigg(\frac{\qbi}{g} \left(v+u f+\frac{\partial f}{\partial\qbat}+\frac{\pbat}{\qbi}\frac{\partial f}{\partial\pbi}\right)
%     \\&
    +\!\frac{({\qbat'})^2}{2\qbi}\!\left({u}\!+\!\frac{1}{g}\!\left(\frac{\partial g}{\partial\qbat}\!+\!\frac{\pbat}{\qbi} \frac{\partial g}{\partial\pbi}\right)\!\!\right)-\frac{{\qbat'}{\qbi'}}{\qbi^2}\!+\!\frac{\qbat''}{\qbi}\Bigg).
\end{align}
{The Poisson bracket of this constraint with itself is now given by \eqref{hhbracketfv}, with}
\begin{align}
    \Gamma_3=F\diff+ \Gamma_4\ham'+\Gamma_5\diff\ham+\Gamma_6\ham ,
\end{align}
{and thus $\Gamma_3$ is zero on the constraint surface.}

Hence, in order to fulfill condition {\it (iii)} of having the
canonical form of the algebra \eqref{eq.hdageneral}, which will allow us to
interpret the Hamiltonian constraint as the generator
of normal deformations, we need to further restrict {its form \eqref{eq.hamU}}.

First, one can check that the coefficients $\Gamma_4$ and $\Gamma_5$ above,
\begin{align}
    \Gamma_4&=-\frac{1}{\qbi}\frac{\partial b_1}{\partial\pbat},   \\
    \Gamma_5&=\frac{1}{\qbi^2}\left(\frac{2}{b_1}\frac{\partial b_1}{\partial\pbat}\frac{\partial b_1}{\partial\pbi}-\frac{\partial^2 b_1}{\partial\pbat\partial\pbi}\right),
\end{align}
vanish if $b_{1}$ is independent of $\pbat$, i.e., $b_{1}=c_1(\pbi,\qbat)$.
Second, implementing this last condition on $\Gamma_6$, we find
\begin{align}
    \Gamma_6=\frac{\qbat'}{\qbi^2}\left(\frac{\partial c_1}{\partial\pbi}-c_1\frac{\partial \log(g)}{\partial\pbi}\right)
\end{align}
and thus $\Gamma_6$ vanishes when ${c_1}(\pbi,\qbat)=\widetilde{\mathfrak{g}}(\qbat)g(\pbi,\qbat)$,
for a generic function $\widetilde{\mathfrak{g}}$. Finally, when we insert these conditions in $F$, we find
\begin{align}\label{structurefunctionpolhamimproved}
    F=\frac{{\widetilde{\mathfrak{g}}^2g^2}}{\qbi^2}\frac{\partial}{\partial\pbi}\left[\frac{1}{g}\frac{\partial f}{\partial\pbi}+\frac{1}{2g}\left(\frac{\qbat'}{\qbi}\right)^{\!2}\frac{\partial g}{\partial\pbi}\right].
\end{align}
In this way, the most general Hamiltonian constraint of the form \eqref{eq.hamqpvac},
that follows the canonical form of the hypersurface deformation algebra \eqref{eq.hdageneral}
with the diffeomorphism constraint \eqref{eq.diff}, is given by
\begin{align}\label{hamSO3vacmod}
    &\ham=\widetilde{\mathfrak{g}}\Bigg({\pbat}\frac{\partial f}{\partial\pbi}+{\qbi}\!\left(v +f{u}+\frac{\partial f}{\partial\qbat}\right)-\frac{{\qbat'}{\qbi'}}{\qbi^2}g
    +\frac{(\qbat')^2}{2\qbi}\left({g{u}}+\frac{\partial g}{\partial\qbat}+\frac{\pbat}{\qbi}\frac{\partial g}{\partial\pbi}\right)+\frac{\qbat''}{\qbi}g\Bigg),
\end{align}
with the structure function \eqref{structurefunctionpolhamimproved}.

In summary, the implementation of the condition {\it (iii)} performed in this subsection
has completely fixed the functional dependence of the constraint \eqref{hamSO3vacmod}
on $E^\varphi$ and $K_x$, while there is still some freedom left on its dependence on
$E^x$ and $K_\varphi$. This freedom is encoded in
the five free functions $f(\pbi,\qbat)$, $g(\pbi,\qbat)$, $u(\qbat)$, $v(\qbat)$,
and $\widetilde{\mathfrak{g}}(\qbat)$.

\subsection{Spacetime embeddability}\label{sec:embeddability}

In order to complete our construction,
in this subsection we will implement the condition {\it {(iv)}} about the spacetime embeddability
of the theory, as given explicitly by Eq. \eqref{eq.covarianceoneoverf}. This relation
ensures that the inverse of the structure function $F$
qualifies as a radial-radial component of the metric.
However, it turns out that Eq.~\eqref{eq.covarianceoneoverf} is not generically satisfied by $F$ as given in \eqref{structurefunctionpolhamimproved}. Therefore, this condition will further restrict the form of the free
functions in the Hamiltonian \eqref{hamSO3vacmod}.

Just in the same way as when solving for anomalies in the previous subsection,
the fact that the functions $f$, $g$, $u$, $v$, and $\widetilde{\mathfrak{g}}$ do not depend on radial derivatives
of the variables allows us to find independent relations for the free functions.
In fact, one can check that the transformation property \eqref{eq.covarianceoneoverf}
is obeyed by the structure function \eqref{structurefunctionpolhamimproved} only if
the following two equations hold,
\begin{subequations}
\begin{align}
g^2 \frac{\partial^3 f}{\partial \pbi^3}-\left(2 {g} \frac{\partial^2 g}{\partial \pbi^2}-\left(\frac{\partial g}{\partial \pbi}\right)^{\!2}\right) \frac{\partial f}{\partial \pbi}&=0,\\
g^2 \frac{\partial^3 g}{\partial \pbi^3}-\left(2g\frac{\partial^2 g}{\partial \pbi^2}-\left(\frac{\partial g}{\partial \pbi}\right)^{\!2}\right)\frac{\partial g}{\partial \pbi}&=0.
\end{align}
\end{subequations}
This system of partial differential equations only includes derivatives with respect to $K_\varphi$,
and the second equation is uncoupled to $f$. Therefore, it is easy to obtain the general solution,
which can be written as follows,
\begin{subequations}\label{fgrule}
\begin{align}
    f&=\left(\frac{A}{\omega^2}\sin^2\big(\omega\pbi+\varphi_f\big)+\chi\right)A_g,\\
    g&=-\frac{A_g}{2}\cos^2\big(\omega\pbi+\varphi_f+\varphi\big),
\end{align}
\end{subequations}
where the six integration functions, $A$, $A_g$, $\varphi_f$, $\varphi$, $\omega$, and $\chi$ depend solely on $\qbat$. {Note that, remarkably, the implementation of the spacetime embeddability condition has completely
fixed the dependence of the Hamiltonian constraint on the variable $K_\varphi$ and only free functions
of $E^x$ survive.}

{We want to point out that} since $f$ and $g$ must be real,
the functions $A$, $A_g$,
and $\chi$ must also be real. However, the integration functions $\omega$, $\varphi_f$, and $\varphi$ can be
either real or complex.
{Besides,} the limit $\omega\to 0$
(with $\varphi_f/\omega\to \phi$, where $\phi=\phi(E^x)$) is well behaved and defines a particular
solution of the family \eqref{fgrule}, with $g$ being independent of $K_\varphi$ and $f$ being quadratic
in $K_\varphi$, i.e., $g=g_0(\qbat)$ and $ f(\pbi,\qbat)=f_0(\qbat)+f_1(\qbat)\pbi+f_2(\qbat)\pbi^2$.
Finally, the condition {\it {(ii)}} implies that $A_g$ cannot be identically vanishing and that $\varphi_f+\varphi\neq\pi/2$ when $\omega=0$.

At this point we have been able to implement the four conditions $({\it i})-({\it iv})$, and
replacing the form \eqref{fgrule} {into \eqref{hamSO3vacmod}} will provide the most general Hamiltonian constraint
we were seeking. Nonetheless, it is possible to see that there is still some redundancy, and the number
of free functions in the Hamiltonian constraint can be reduced by performing certain redefinitions. More precisely,
we get rid of $\varphi_f$ through the canonical transformation $\pbi\to\pbi-\varphi_f/\omega$ and $\pbat\to\pbat-\qbi\frac{\partial(\varphi_f/\omega)}{\partial\qbat}$, which leaves invariant the diffeomorphism constraint. 
Further, we set $\widetilde{\mathfrak{g}}=-\mathfrak{g}/A_g$ and $A_g=\exp\left[\int \big(B-u\big) d\qbat\right]$, with $B=B(\qbat)$. We also introduce yet another function $V:= - (v +B\chi +\frac{\partial\chi}{\partial\qbat})\exp\left[\int \big(u-B\big) d\qbat\right]$, so that all the ``potential'' terms are 
gathered in a unique function.
Finally, it is also convenient to define $W(\qbat):=\exp\big[\int B \,d\qbat\big]$.

{
\subsection{Result}\label{sec:result}
}

Taking all the above into account, the most general Hamiltonian constraint that obeys conditions
\textit{{(i)}}-\textit{{(iv)}}, and thus covariantly defines the metric \eqref{eq.metric}, reads 
\begin{align}\label{hamSO3vacmod+}
    &\ham={\mathfrak{g}}\Bigg(\qbi V-{\qbi}\frac{A}{\omega^2}\sin^2\big(\omega\pbi\big)\frac{\partial }{\partial\qbat}\left[\log\left(\!\frac{A\,W}{\omega^2}\!\right)\right] +\frac{1}{2}\left(\frac{\qbat''}{\qbi}-\frac{{\qbat'}{\qbi'}}{\qbi^2}+\frac{\partial \log(W)}{\partial\qbat}\frac{(\qbat')^2}{2\qbi}\right)\cos^2\big(\omega\pbi+\varphi\big) \nonumber\\[6pt]
    & -\left({\pbat}+\pbi\qbi\frac{\partial\log(\omega)}{\partial\qbat} \right)\frac{A}{\omega}\sin\big(2\omega\pbi\big)
    -\left(\omega{\pbat}+\pbi\qbi\frac{\partial\omega}{\partial\qbat}+\qbi\frac{\partial\varphi}{\partial\qbat}\right)\left(\frac{\qbat'}{2\qbi}\right)^{\!2}\sin\big(2(\omega\pbi+\varphi)\big) \Bigg),
\end{align}
where ${\mathfrak{g}}={\mathfrak{g}}(E^x)$, $\omega=\omega(E^x)$, $\varphi=\varphi(E^x)$, $A=A(E^x)$, $V=V(E^x)$, and $W=W(E^x)$ are free functions of the variable $\qbat$.
This result is unique up to canonical transformations that respect the form of the diffeomorphism constraint \eqref{eq.diff} and do not include derivative terms.
As explained in Appendix~\ref{app.bd}, this Hamiltonian constraint is equivalent
to the one recently presented in Ref. \cite{Bojowald:2023xat}.

Recall that, as long as the Hamiltonian is real,
the arguments of the trigonometric functions may be complex,
changing to hyperbolic solutions, and that the limit $\omega\to0$ is well defined.
In fact, it is easy to see that such limit, along with the choice of
functions $\mathfrak{g}=\sqrt{\erad}$, $V=-1/(2\erad)$, $A=1$, $W=\sqrt{\erad}$, and $\varphi=0$,
renders \eqref{hamSO3vacmod+} into \eqref{eq.hamGR}, and thus corresponds to the particular case
of vacuum GR. {
In fact, the choice {$V=-1/(2\erad)+{\Lambda/2}$}, with a constant $\Lambda$,
reproduces a cosmological constant term, which
should obviously be allowed by covariance.
}

The structure function $F$ that appears in the bracket \eqref{eq.hhqp}, considering the above Hamiltonian constraint
\eqref{hamSO3vacmod+},
reads
\begin{align}\label{eq.F}
    F&=\frac{F_s}{\qbi^2},\quad\mathrm{with}\quad
 F_s:={\mathfrak g}^2\cos(\omega K_\varphi+\varphi)\left(
 A\cos(\omega K_\varphi-\varphi)+\left(\frac{{E^x}'}{2 E^\varphi}\right)^2\omega^2\cos(\omega K_\varphi+\varphi)
 \right).
\end{align}
At this point, it is very convenient to define the function
\begin{align}\label{eq.defsm&G}
    \mass&:=\left(\frac{\qbat'}{2\qbi}\right)^2\cos^2\big(\omega\pbi+\varphi\big)-A\frac{\sin^2\big(\omega\pbi\big)}{\omega^2},
\end{align}
which can be shown to be a spacetime scalar, as its transformation {(on-shell)} is given by
\begin{align}
    \left\{\mass,H\big[ \epsilon^t N\big]+D\big[\epsilon^tN^x+\epsilon^x\big]\right\}\approx\epsilon^t\dot{\mass} +\epsilon^x \mass'.
\end{align}
This function can be used to reexpress the Hamiltonian constraint in the following compact form,
\begin{align}\label{hamSO3vacmodm&G}
\!\!\!\ham  ={\mathfrak{g}}\frac{\qbi}{\qbat'}\Bigg(\!\!\left( {V}\!+\frac{\partial\log(W)}{\partial\qbat}\mass\right)\qbat'+\mass'
     -\frac{\diff}{\qbi}\frac{\partial \mass}{\partial\pbi} 
    \Bigg).
\end{align}
From here, one can deduce that on the constraint surface ($\ham\approx0$ and $\diff\approx0$)
$M$ is a function of $\qbat$ only, and it is explicitly given by
\begin{align}\label{Monshell}
    M\approx  -\left(W(\qbat)+\int V(\qbat)\,d\qbat\right),
\end{align}
up to an integration constant. In terms of this function, $F_s$ also takes a very simple form,
\begin{align}
\label{eq.Fs}
F_s=   {\mathfrak{g}}^2\Big({A}\cos^2(\varphi)+\omega^2\mass\Big).
\end{align}
In turn, this shows that $F_s$ is also a spacetime scalar and, on-shell, it is a function
of $\qbat$ only. 

{Let us finish this section with some remarks.
It is very interesting to note that the covariance requirement --i.e., conditions {\it (iii)}
and {\it (iv)}-- has severely restricted the functional dependence of the Hamiltonian constraint
\eqref{hamSO3vacmod+}
on the different variables. In particular, its dependence on $E^\varphi$, $K_x$, and $K_\varphi$
is completely fixed, and only free functions of $E^x$ are allowed. More precisely,
the variable $K_x$ only appears linearly, while the dependence on $E^\varphi$ is a bit more involved,
but, apart from the derivative term ${E^\varphi}'$, it only appears either linearly
or with certain inverse powers. Concerning $K_\varphi$, it appears in several terms, both
inside and outside the argument of trigonometric functions (which, as commented above,
can also be hyperbolic). These trigonometric functions may be of special relevance
in the context of effective models of loop quantum gravity, since, motivated by
the holonomy variables that are used in this theory for the quantization,
the building of effective models has been based on the so-called polymerization,
which consists on replacing the curvature degrees of freedom, like $K_\varphi$,
by certain trigonometric function.
In this respect, we would like to stress that the trigonometric functions that
appear in the Hamiltonian constraint \eqref{hamSO3vacmod+} are a direct consequence of the
implementation of conditions {\it (iii)} and {\it (iv)}.
}

In summary, we have obtained the family of Hamiltonian constraints \eqref{hamSO3vacmod+}, which generalize
GR in spherical symmetry, though keeping the good covariance properties,
which will allow us to provide a consistent spacetime metric.
As will be explained below in more detail, in general,
these models are not equivalent to GR and they lead to different spacetime geometries,
thus the uniqueness results \cite{Kuchar:1974es,Hojman:1976vp} do not apply here.
This has already been shown in Refs. \cite{Alonso-Bardaji:2021yls,Alonso-Bardaji:2022ear,Alonso-Bardaji:2023niu}, where
particular cases of the Hamiltonian constraint \eqref{hamSO3vacmod+} were studied in detail.

\section{Structure of the spacetime}\label{sec.st}

In this section we provide the spacetime geometry {defined} by the models constructed in
the previous section, and analyze its main features. This section is divided in four subsections. In Sec. \ref{sec:metric} the full four-dimensional
spacetime metric is constructed. In Sec. \ref{sec:curvature} we present the curvature invariants. In Sec. \ref{sec:killing} we show that all the spacetimes {under consideration have}
a Killing vector field in the ${\cal M}^2$ {sector}. Finally, in Sec. \ref{sec:discussion}
we discuss the general properties and structure of these geometries.

\subsection{The spacetime metric}\label{sec:metric}

By construction, there are no propagating degrees of freedom in this theory
since there are two conjugate couple of variables, $(E^x,K_x)$ and $(E^\varphi,K_\varphi)$,
and two first-class constraints.
The Hamiltonian constraint \eqref{hamSO3vacmod+}, along with the diffeomorphism
constraint \eqref{eq.diff}, and the Lagrange multipliers
($N$, $N^x$) define the Hamiltonian $H[N]+D[N^x]$, which encodes
the dynamics of the system. As usual, one can obtain the equations
of motion for the different variables through the Poisson brackets,
\begin{align}
\dot{E}^x=\{E^x,H[N]+D[N^x]\},\label{eq.exdot}\\
\dot{K}_x=\{K_x,H[N]+D[N^x]\},\\
\dot{E}^\varphi=\{E^\varphi,H[N]+D[N^x]\},\\
\dot{K}_\varphi=\{K_\varphi,H[N]+D[N^x]\}.
\end{align}
The solution to these equations in a given gauge will provide the metric
tensor of the two-dimensional manifold ${\cal M}^2$
in certain coordinate system:
\begin{equation}
g_{AB}dy^A dy^B=\s N^2 dt^2+\frac{1}{|F|} (dx+N^x dt)^2,        
\end{equation}
with $F$ as defined in \eqref{eq.F}. Note that, in principle, the sign of $F$ is not fixed,
and thus the signature of this metric is encoded in ${\s}:=-{\rm sgn}(F)$.
Due to the construction performed above,
a change of gauge will simply correspond to a change of coordinates,
and thus this {(two-dimensional)} line element is covariantly defined.

Now, in order to provide a complete four-dimensional geometric picture
of the spacetime, this line element must be extended.
Our aim is to describe spherically symmetric spacetimes, and thus, following
the discussion of Sec. \ref{sec:gr}, we will
assume that the four-dimensional manifold ${\cal M}$ is a warped product
 ${\cal M}={\cal M}^2\times S^2$, with line element,
\begin{equation}\label{eq:4metric}
 ds^2= g_{AB} dy^A dy^B+r^2 d\Omega^2,
\end{equation}
where $r$ is the function that provides the area of the spheres.
Due to the symmetry reduction, the Hamiltonian only contains information about the
geometry of the manifold ${\cal M}^2$, but it knows nothing about $r$.
The only requirement for this function is that it should be a spacetime scalar.
Since, among our basic variables, $E^x$ is the only spacetime scalar,
one can assume $r=r(E^x)$ in general, and thus $r$ will be (another) independent
free function of the model. For the particular case of GR its value is given by $r=\sqrt{E^x}$.

With the full four dimensional metric \eqref{eq:4metric} at hand, we can move on to study
the geometry of the spacetime.
Even if our construction has severely restricted the freedom of the model,
there are still seven free functions ($A$, $V$, $W$, $\mathfrak{g}$, $\omega$, $\varphi$, and $r$)
of the variable $E^x$, and the specific properties of the spacetime will
depend on their precise form.
However, {we will be able to conclude} some relevant features
quite generically.

\subsection{Curvature}\label{sec:curvature}

For a spherically symmetric metric of the form \eqref{eq:4metric},
the gradient of the area-radius function $v_A:=\nabla_A r$,
where $\nabla$ is the covariant derivative associated with the two-dimensional
metric $g_{AB}$, contains key physical information of the spacetime.
In any coordinate system, the form of this vector
is explicitly given by,
\begin{equation}\label{eq.defv}
 v_Ady^A=\frac{dr}{d E^x}[\dot{E}^x dt+(E^x)'dx],
\end{equation}
where, making use of the equation of motion \eqref{eq.exdot}, 
\begin{equation}
\dot{E}^x=N^x {E^x}'+\frac{2N\mathfrak{g}}{\omega \cos(\omega K_\varphi+\varphi)}
[( A +\omega^2 M ) \sin(\omega K_\varphi)\cos(\varphi)+\omega^2 M \cos(\omega K_\varphi)\sin(\varphi)].
\end{equation}

In particular, the norm of $v_A$ will be very relevant in the analysis of the structure of the spacetime,
since the spheres of constant $t$ and $x$ will be trapped or not depending on its sign ({because the mean curvature vector of those spheres is $2v^A/r$}). Therefore,
contracting the indices with the metric \eqref{eq:4metric}, we define
\begin{align}\label{defH}
    \hs:=v_A v^A \approx  -4 \,\s\, \left(\frac{d r}{d \qbat}\right)^2\mathfrak{g}^2 \mass\Big(A+\omega^2\mass\Big).
\end{align}
The sign of $\hs$ is thus completely characterized by
$A$, $\omega$, and $M$. For the Riemannian ($\s=1$) case $\hs$ is always positive
or vanishing if $v_A=0$. For the Lorentzian ($\s=-1$) case,
in the regions where $\hs$ is negative (positive) the corresponding spheres are trapped (nontrapped),
while
the hypersurfaces where $\hs=0$ are either
marginally trapped (if $v_A\neq 0$) or minimal
(if $v_A=0$).

In addition, all the information regarding the spacetime curvature is encoded in the norm $\hs$ of the vector $v^A$, the trace of its gradient $\nabla_A v^A$, and the Ricci scalar $R$ of the two dimensional metric $g_{AB}$. More precisely, there are only two independent spacetime curvature scalars,
namely the four-dimensional Ricci scalar,
\begin{equation}
{}^{(4)}R=R+\frac{2}{r^2}\big(1-\hs\big)-\frac{4}{r}\nabla_Av^A,
\end{equation}\label{eq.ricci}
and the scalar,
\begin{equation}\label{eq.curvaturescalar}
{\cal U}=-\frac{1}{6}\left(R+\frac{2}{r^2}\big(1-\hs\big)+\frac{2}{r}\nabla_Av^A\right),
\end{equation}
which provides all nonzero components of the Weyl tensor,
\begin{align}
{}^{(4)}W_{ABCD}&=(g_{AD}g_{BC}-g_{AC}g_{BD})\,{\cal U},\\
{}^{(4)}W_{A\phi B\phi}&={}^{(4)}W_{A\theta B\theta}\sin^2\theta=\frac{r^2}{2}\sin^2\theta\,g_{AB}\,{\cal U},\\
{}^{(4)}W_{\theta\phi\phi\theta}&=r^4\sin^2\theta\,{\cal U},
\end{align}
and permutations, where $\theta$ and $\phi$ are the angular coordinates, {with $d\Omega^2=d\theta^2+\sin^2\theta d\phi^2$.}

Depending on the specific form of the free functions in the Hamiltonian constraint \eqref{hamSO3vacmod+},
these curvature invariants might have a very different behavior and, in particular,
might diverge at one or several values of $E^x$.
In the particular case of vacuum GR, a divergence appears at $E^x\to 0$ ($r\to 0$),
which signals the classical singularity. In a modified theory, this divergence
might still be present, and even new ones appear at different values of $r$.
However, there might also be singularity-free spacetimes. This was indeed the case in Refs. \cite{Alonso-Bardaji:2023niu,Alonso-Bardaji:2022ear,Alonso-Bardaji:2021yls}, where the domain of $E^x$ was constrained to certain regions with no curvature divergences (in contrast to GR, where the domain of
$\erad$ is the whole positive real line).

\subsection{Killing vector fields}\label{sec:killing}

Since the spacetime \eqref{eq:4metric} is spherically symmetric, there are three independent Killing vector fields on $S^2$. However, not all spherically symmetric spacetimes contain Killing fields on the sector $\mathcal{M}^2$. In this section, we will show that such a Killing field exists for all the theories defined by the Hamiltonian constraint
\eqref{hamSO3vacmod+}. This is a nontrivial result, though one would intuitively expect that this is the case,
due to the fact that, by construction, there are no propagating degrees of freedom in the dynamics
described by this family of deformed Hamiltonians.

The Killing equation reads
\begin{equation}
 \xi_{\mu;\nu}+\xi_{\nu;\mu}=0,
\end{equation}
where $\mu,\nu=0,1,2,3$ are four-dimensional indices and the semicolon $;$
stands for the covariant derivative associated with the four-dimensional metric \eqref{eq:4metric}.
Assuming that the angular components of the vector are vanishing, that is $\xi_\mu dx^\mu=\xi_A dx^A$,
it is easy to see that the above equation can be decomposed in the following relations:
\begin{align}
 \xi_{A, \theta}=\xi_{A, \phi}=0,\label{eq:killingang}\\
  \xi^A v_A=0,\label{eq:killingorthogonal}\\
 \nabla_{\!A}\xi_{B}+\nabla_{\!B}\xi_{A}=0, \label{eq:killingM2}
\end{align}
where $\nabla$, as already defined above, is the covariant derivative compatible with the two-dimensional
metric $g_{AB}$ on ${\cal M}^2$. It is clear that relations \eqref{eq:killingang} simply imply that $\xi_A$ is independent
of the angular coordinates, while \eqref{eq:killingorthogonal} requests the Killing vector to be orthogonal
to $v_A$. Therefore, in all the points where $v_{A}$ is not vanishing,
the most general vector that obeys relations \eqref{eq:killingang}--\eqref{eq:killingorthogonal} is given by
\begin{equation}
 \xi^A=h \, u^A,\label{eq:killingform}
\end{equation}
where $h$ is a scalar function on ${\cal M}^2$ and the vector $u^A:=\epsilon^{AB}v_{B}$,
with $\epsilon^{AB}$ being the covariant Levi-Civita tensor on ${\cal M}^2$
\footnote{That is,
$\epsilon^{AB}=\frac{1}{\sqrt{|g|}}\,\eta^{AB}$, with $\eta^{AB}$ being the antisymmetric symbol with value $\eta^{01}=1$ and $g$ the determinant of $g_{AB}$.
},
defines the orthogonal direction to $v_{A}$.

The remaining equation \eqref{eq:killingM2} can be decomposed in three scalar
equations simply by projecting it along the vector $v^{A}$ and its orthogonal $u^A$\footnote{Note that the metric is given by $g_{AB}=({\s} u_A u_B +v_{A} v_{B}) /\hs$.},
\begin{align}
 v^{A}v^{B}\nabla_{\!A}\xi_{B}=0,\\ v^{A}u^B(\nabla_{\!A}\xi_{B}+\nabla_{\!B}\xi_{A})=0,\\
 \nabla_{\!A}\xi^{A}{}=0.
\end{align}
Replacing the form \eqref{eq:killingform} of the Killing vector field, these three equations take the form,
 \begin{align}
 h\, u^{A}v^B\nabla_A v_B=0, \label{eq:killing1}\\
\hs\, v^{A} \nabla_{\!A}h  - h\big(\hs \nabla_{\!A}v^A-2v^Av^B \nabla_{\!A}v_B\big) =0,\label{eq:killing3}\\
 u^A \nabla_{\!A}h=0.\label{eq:killing4}
 \end{align}
The last relation \eqref{eq:killing4} implies that $\nabla_{\!A}h$ must be proportional to $v_{A}$,
and thus $h$ must be a function of $r$ only, i.e., $h=h(r)$. Then, \eqref{eq:killing3} is a first-order linear ordinary differential
equation for $h(r)$, which can be rewritten as follows,
\begin{equation}\label{eq.killingdifeq}
\hs^2 \frac{dh(r)}{dr}  {-}  h(r)(\hs \nabla_{\!A}v^{A}-2v^{A}v^{B} \nabla_{\!A}v_B)=0.
\end{equation}
If $\hs$ is nonvanishing,
this can be further simplified writing the term inside brackets as a total derivative,
\begin{equation}
\hs^2 \left(\frac{d h(r)}{dr}  -  h(r)\nabla_{\!A}\left(\frac{v^{A}}{\hs}\right)\right)
=0.
\end{equation}
Then, for $\hs\neq 0$,
 the solution to this
equation reads
\begin{equation}\label{eq:hsolution}
 h(r)=\exp\left[\int dr \,
 \nabla_{\!A}\left(\frac{v^{A}}{\hs}\right)  \right],
\end{equation}
where a global integration constant has been fixed without loss of generality
(it only amounts to a constant rescaling of the Killing field).
{
Making use of the equations of motion and the weak equality \eqref{Monshell},
it is possible to perform this integral and obtain the following simple expression, 
\begin{equation}\label{eq:hsolutionexplicit+}
h^2\approx\frac{W^2}{4\left(\frac{dr}{dE^x}\right)^2{\mathfrak g}^2{|F_s|}},
\end{equation}
}%
which explicitly provides $h$ in terms of the free functions of the model
for $\hs\neq 0$. In the case $\hs=0$, which implies a lightlike vector $v_A$
(recall that for this derivation we are assuming $v_A\neq 0$),
equation \eqref{eq:killing3} requires either
$v^A\nabla_A\hs=0$ or $h=0$. Therefore, at points where $\hs=0$
and $v^A\nabla_A\hs\neq 0$ the Killing vector field vanishes ($h=0$). {For points
with $\hs=0=v^A\nabla_A\hs$ the form of $h$ is not restricted by Eq. \eqref{eq:killing3},
and thus in general the Killing field will be lightlike there
(unless continuity, or some other condition, on the function
$h$ requires it to be vanishing at this point).}

The only nontrivial equation left is \eqref{eq:killing1}. This equation does not restrict
further the form of $h$, it rather imposes a necessary condition on the spacetime geometry
for the existence of the Killing vector field. It can be rewritten as,
\begin{equation}
u^A\nabla_{\!A} \hs=0. \label{eq:killingcondition}
\end{equation}
That is, the gradient of $\hs$ must be orthogonal to $u^A$, and thus proportional to $v_{A}$.
This implies that the Killing vector will exist if $\hs$ is a function of $r$ only.
As can be seen in Eq.~\eqref{defH}, this is indeed the case for the family of Hamiltonians derived
here, as long as the function $r=r(E^x)$ is invertible.
This completes the proof of the existence of the Killing field.

In this way, we have shown that, as long as $v_A\neq 0$,
all the spacetimes under consideration have a Killing vector field $\xi^A$ in the ${\cal M}^2$ sector.
This Killing field is orthogonal to $v_A$ and takes the form \eqref{eq:killingform},
with the function $h$ given by \eqref{eq:hsolutionexplicit+} for $\hs\neq 0$.
For hypersurfaces $r=a$ where the vector $v_A\neq 0$ is lightlike, and thus $\hs(a)=0$,
{
there are two possibilities: if $v^A\nabla_A\hs\neq 0$ the Killing vector $\xi_A$
is also require to vanish there $(h=0)$, while if $v^A\nabla_A\hs= 0$ the function $h$
is free and in general $\xi_A$ will be lightlike.}
On the other hand,
at minimal hypersurfaces where $v_A=0$, the above analysis does not apply,
Eq. \eqref{eq:killingorthogonal} is automatically satisfied, and the existence of the Killing field will depend
on condition \eqref{eq:killingM2} being fulfilled. However, if the vector $v_A$ vanishes only locally at $r=a$,
and thus $\nabla_A v_B\neq 0$ there, and the spacetime {is smooth around that point}
\footnote{{By smoothness of a region of the spacetime we will imply that all the fields are locally analytic.}},
the Killing vector will be defined at $r=a$ by continuity.

As a side remark, in the appendix we present the construction and the conditions for the existence of the Killing
field in terms of the Einstein tensor. In this way,
one can see how the mentioned conditions are automatically
fulfilled for vacuum general relativity, which makes contact with the usual Birkhoff's theorem.

\subsection{General features of the spacetimes}\label{sec:discussion}

The dynamics given by the family of Hamiltonian constraints \eqref{hamSO3vacmod+}
will define a spacetime with metric \eqref{eq:4metric}.
In the particular case of GR, the scalar part of the structure function reads $F_s=E_x$, which
is positive for all positive values of $E^x$, and the area-radius function is $r=\sqrt{E^x}$.
This construction defines a Lorentzian manifold (the Schwarzschild spacetime),
with the area of spherical orbits taking all values between zero and infinity,
and a singularity located at $r\to 0$, where the curvature {scalar \eqref{eq.curvaturescalar}
diverges.}

However, in the general case, $F_s=F_s(E^x)$ will not be defined for all positive values of $E^x$,
and the domain of $E^x$ will be restricted to certain finite or infinite domains.
Each {disjoint} domain will define an independent spacetime, where,
once the function $r=r(E^x)$ is fixed, the area of the spheres
will correspondingly be restricted. The signature of the metric at each value
of $E^x$ will be given by the sign of $F$ there.
As long as $F_s(E^x)\neq 0$, the signs of $F_s$ and $F$ coincide, and thus $F_s$
encodes also the signature. However, roots $z$ of $F_s(E^x=z)=0$ need to be
analyzed carefully. In general, these roots will imply a boundary of a given
Lorentzian or Riemannian region. 
Depending on the behavior of the dynamics there, these points may or may not be part
of the spacetime manifold. Also, they may or may not be traversable, in the sense that there
might exist an open subset of $E^x$ around $E^x=z$, where $F_s$ is well defined.
Although such traversability does not necessarily imply a signature change since
the sign of $F_s$ may be the same for the whole open subset.
With all the free functions of the model, the amount of possible cases
is very wide, so
a case by case analysis should be performed once the form of the different functions
has been fixed.

Nonetheless, despite the freedom that the model still encodes,
there are several general properties of the spacetime that we can draw.
In these spherical spacetimes, there are two key vectors: the gradient of the
area-radius function $v_A$ \eqref{eq.defv}, with its norm $\hs$ given in \eqref{defH},
and the Killing vector field $\xi_A$. As long as $v_A\neq 0$, from Eq.~\eqref{eq:killingform}
it is straightforward to see that
these vectors are orthogonal $\xi_Av^A=0$, and that the norm of the Killing field is given by
\begin{equation}\label{killingmodule}
 G:=\xi_A \xi^A=\sigma h^2 \hs.
\end{equation}
For a Riemannian spacetime $(\sigma=1)$ both $G$ and $\hs$ have a positive sign,
while for a Lorentzian spacetime $(\sigma=-1)$ they have opposite signs.
This fact defines two generic regions for these Lorentzian spacetimes:
\begin{itemize}
 \item $G<0$ and $\hs>0$: static nontrapped regions, where $\xi_A$ is timelike
 and $v_A$ spacelike, like the exterior of the Schwarzschild black hole. 
  \item $G>0$ and $\hs<0$: homogeneous trapped regions, where $\xi_A$ is spacelike
 and $v_A$ timelike. Depending on the direction of $v_A$, in these regions the spheres
 can be trapped to the future, like in
the interior of the Schwarzschild black hole, or trapped to the past (antitrapped),
like in the interior of the Schwarzschild white hole.
\end{itemize}

Roots of the scalar functions $G=G(E^x)$ and $\hs=\hs(E^x)$ define, in general, three-dimensional hypersurfaces
on the manifold, where $E^x$ is constant (and, once an invertible relation $r=r(E^x)$ has been fixed,
will imply a constant $r$).
In the Riemannian case, a root of these scalar functions imply
that the corresponding vector ($\xi^A$ and $v^A$, respectively) vanishes there. However, in
the Lorentzian case, the vanishing of $G=G(E^x)$ or $\hs=\hs(E^x)$ may imply two different
things: either the corresponding vector vanishes
or it is lightlike. Clearly, if they imply a change of sign in a smooth region of the spacetime,
these roots will happen simultaneously (i.e., $\hs(z)=0$ and  $G(z)=0$),
and thus define a boundary $E^x=z$ between a static and a homogeneous region.
But in general there will be roots of $\hs$, which will not be roots of $G$.

Note that relation \eqref{eq:killingform}, and thus \eqref{killingmodule},
is valid for $v_A\neq 0$. Therefore, a lightlike $v_A$ implies either a vanishing (if $h=0$)
or a lightlike (if $h\neq 0$) Killing field $\xi_A$.
But, in general, at hypersurfaces where $v_A=0$ (and thus $\hs=0$),
$\xi^A$ does not need to be lightlike or vanishing (and thus $G\neq 0$).
Nonetheless, if $v_A$ vanishes only locally at a given hypersurface,
and the spacetime is smooth around that hypersurface,
relation \eqref{killingmodule} should also be obeyed there.
Hence, under the assumption of smoothness,
$v_A=0$ also implies a lightlike (if $h\neq 0$) or vanishing (if $h=0$)
$\xi^A$, unless $h$ diverges.
Interestingly, from \eqref{eq:hsolutionexplicit+}, we see that
this latter is indeed the case if $F_s$ also vanishes there,
while $\frac{W^2}{\mathfrak{g}^2 \left(\frac{dr}{dE^x}\right)^2}\neq 0$.
Therefore, in a smooth region of the spacetime, there can be a hypersurface
where $v_A=0$, but with a neither lightlike nor vanishing $\xi^A$
(i.e., $G\neq 0$), if $F_s$ vanishes and $\frac{W^2}{\mathfrak{g}^2 \left(\frac{dr}{dE^x}\right)^2}\neq 0$ there.
However, the modulus of the Killing field $G$ to be finite,
it is necessary that $\hs$ vanishes there too.

Let us thus check the possible simultaneity in the roots of $F_s$
and $\hs$. From their expressions {\eqref{eq.Fs}} and \eqref{defH}, one can write in general
\begin{equation}\label{eq.HF}
 {\mathfrak g}^2\omega^2 \hs= -4\sigma \left(\frac{dr}{dE^x}\right)^2(F_s-{\mathfrak g}^2 A \cos^2\varphi)(F_s+{\mathfrak g}^2 A \sin^2\varphi),
\end{equation}
which, for a vanishing $F_s$, leads to the relation
\begin{equation}\label{eq.hsFszero}
 {\mathfrak g}^2\omega^2 \hs\big{|}_{F_s=0}= 4\sigma \left(\frac{dr}{dE^x}\right)^2{\mathfrak g}^2 A^2\sin^2(2\varphi).
\end{equation}
Therefore, if ${\mathfrak g}\, \omega\neq 0$
at this hypersurface, either $\hs=0$ or ${\rm sgn}(\hs)=\sigma$.
That is, in a Lorentzian spacetime a hypersurface where $F_s=0$ and ${\mathfrak g}\, \omega\neq 0$
is either embedded in a homogeneous region (where $\hs<0$) or $\hs=0$ there.
This latter is the case, in particular, if $\sin(2\varphi)=0$, either locally at that hypersurface
or exactly for the whole spacetime. Since an exact vanishing of $\sin(2\varphi)$ defines a subfamily of models with certain interesting features, let us analyze them in a bit more detail.

\subsubsection*{The subfamily of models with $\varphi=n\pi/2$}

A particularly interesting subfamily of models, for which a root of $F_s$ always implies
a vanishing $\hs$, given that ${\mathfrak g}\, \omega\neq 0$, corresponds to the case
when the free function $\varphi=\varphi(E^x)$ reads $\varphi=n\pi/2$, for any integer $n$
\footnote{Note that, for such values of $\varphi$, the Hamiltonian \eqref{hamSO3vacmod+} to be real
$\omega$ must be either real or purely imaginary.}.
But, before analyzing the behavior of the roots of $F_s$, let us first note that
the constant form of the function $\varphi$ under consideration
introduces certain symmetries in the system.
In particular, from Eq.~\eqref{eq.F} we see that, when $\varphi=n\pi/2$,
the two terms inside the parenthesis have the same phase,
and the scalar part of the structure function is then given as
\begin{equation}\label{eqFphi0}
 F_s={\mathfrak g}^2\cos^2\left(\omega K_\varphi+\frac{n\pi}{2}\right)\left(
 (-1)^n A+\left(\frac{{E^x}'}{2 E^\varphi}\right)^2\omega^2
 \right).
\end{equation}
From this expression it is straightforward to conclude that, if $A$ has a definite sign
for all values of $E^x$, the sign of $F_s$ will also be fixed in the following instances
\footnote{Take into account that $\cos^2(i x+n\pi/2)=\cosh^2(x)$ for $n$ even, and $\cos^2(i x+n\pi/2)=-\sinh^2(x)$ for $n$ odd.}:
\begin{itemize}
\item $\omega$ real and $(-1)^nA\geq 0$ for all $E^x\,\Longrightarrow F_s\geq 0\Longrightarrow$ Lorentzian signature.
\item $n$ odd, $\omega$ purely imaginary, and $A\geq 0$ for all $E^x\,\Longrightarrow F_s\geq 0\Longrightarrow$ Lorentzian signature.
\item $n$ even, $\omega$ purely imaginary, and $A\leq 0$ for all $E^x\,\Longrightarrow F_s\leq 0\Longrightarrow$ Riemannian signature.
\end{itemize}
Therefore, these cases can
only describe either Lorentzian or Riemannian spacetimes,
and a signature change is completely excluded.

In addition, from \eqref{eqFphi0}, one can see that $F_s$ can vanish either because the term in parenthesis
vanishes, or because the global factor does. On the one hand, if the term in parenthesis vanishes,
at $F_s=0$ one has that $(-1)^nA/\omega^2=-({E^x}'/(2E^\varphi))^2\leq 0$.
On the other hand, if the global factor vanishes, since it has a definite sign,
then the signature around $F_s=0$ will be given by the term in parenthesis:
for a Riemannian region one then has
$F_s\leq0\Rightarrow (-1)^n A/\omega^2\leq 0$, while for a Lorentzian region
the sign of $(-1)^n A/\omega^2$ is not fixed.

As commented above,
the form of the function $\varphi=n\pi/2$ makes $\sin(2\varphi)$ to be exactly vanishing, and thus the
right-hand side of \eqref{eq.hsFszero} vanishes. Since $\hs=0$,
in principle, the vector $v_A$ might be either lightlike or vanishing at that hypersurface.
Assuming smoothness and making use of relations \eqref{eq:hsolutionexplicit+}, \eqref{killingmodule},
and \eqref{eq.HF}, it is easy to obtain the norm of the Killing field
there,
\begin{equation}
 G\big{|}_{F_s=0}= \frac{\sigma W^2 A}{\mathfrak{g}^2\omega^2}(-1)^{n+1}.
\end{equation}
This expression shows that, if $W^2 A$ is not vanishing at that hypersurface,
the Killing field is either spacelike or timelike, but not lightlike,
which implies that the vector $v_A$ is vanishing there.
In summary, for this particular subfamily of models, under the commented assumptions
(smoothness, $\mathfrak{g}^2\omega^2\neq 0$ and $W^2 A\neq 0$ locally),
the roots of $F_s$ always define minimal hypersurfaces, where $v_A=0$
\footnote{Note that it is immediate to extend this result to a local vanishing of $\sin(2\varphi)$.}.
In a Lorentzian spacetime ($\sigma=-1$) the Killing field is
spacelike (for $(-1)^n A/\omega^2>0$) or timelike (for $(-1)^n A/\omega^2<0$) at that hypersurface.
And, since neither $G$ nor $\hs$ changes sign there, this hypersurface
simply defines a boundary between two homogeneous (if $(-1)^n A/\omega^2>0$)
or between two static (if $(-1)^n A/\omega^2<0$) regions.
For a Riemannian spacetime ($\sigma=1$), as commented above,
$(-1)^n A/\omega^2$ is always negative {at $F_s=0$}, which makes, as expected, $G\big{|}_{F_s=0}$ to be positive.

For a detailed analysis of a particular nontrivial case of this subfamily of models,
which shows some of the features described above,
we direct the reader to Refs.~\cite{Alonso-Bardaji:2021yls,Alonso-Bardaji:2022ear}, where a regular version of
the Schwarzschild black hole was constructed. Such model corresponds to the choice of functions
$\varphi=0$, $\omega$ real, and $A>0$ for all $E^x$, and thus implies
a non-negative $F_s$ with Lorentzian signature.
Furthermore, in that case $F_s$
turns out to be only defined in the interval $r\in[r_0,\infty)$, for some finite $r_0$.
At the hypersurface $r=r_0$ the scalar part of the structure function $F_s$ vanishes, and
it is therefore a minimal hypersurface with $v_A=0$ and a spacelike Killing field. This hypersurface
is a boundary between two homogeneous (a trapped and an antitrapped) regions. This spacetime turns out to
be completely regular and geodesically complete.
Another, more involved, example, but also with a non-negative $F_s$,
is presented in Ref. \cite{Alonso-Bardaji:2023niu}, where the previous
black-hole model is endowed with electric charge and a cosmological constant.

\section{Minimal coupling of matter}\label{sec.matter}

As explained above, the models presented here for vacuum spherical gravity are nondynamical.
In order to describe a dynamical process, like a black hole formation or an explosion,
it is necessary to provide a prescription to couple matter degrees of freedom
to the model in a consistent way.

The model admits matter following the usual minimal-coupling prescription.
Let us, for instance, consider a scalar field with
Lagrangian density
\begin{align}
    \mathcal{L}_m:&=-\frac{1}{2}\sqrt{|g|}\big(g^{\mu\nu}\partial_\mu\psi\partial_\nu\psi+V(\psi)\big).
\end{align}
Using the metric \eqref{eq:4metric}, and assuming that $\psi$ is spherically symmetric, and thus independent of the angular coordinates, we get
\begin{align}
    \mathcal{L}_m &=-\frac{ r^2}{2N\sqrt{|F|}}\Big(\s(\dot{\psi}-N^x\psi')^2+|F|N^2(\psi')^2 +{ N^2V}\Big).
\end{align}
The conjugate momentum of the scalar field is defined as
\begin{align}
    P_\psi:=\frac{\partial\mathcal{L}_m}{\partial\dot{\psi}}=-\frac{\s r^2}{N\sqrt{|F|}}\Big(\dot{\psi}-N^x\psi'\Big),
\end{align}
and can be inverted to write the time derivative
\begin{align}
    \dot{\psi} =-\s N\sqrt{|F|}\frac{P_\psi}{ r^2}+N^x\psi'.
\end{align}
We can then perform a Legendre transformation to obtain
the total matter Hamiltonian,
\begin{align}
    \dot{\psi}P_\psi-\mathcal{L}_m = N\ham_m+N^x\diff_m,
\end{align}
which is a sum of constraints, with
\begin{align}
    \diff_m&:=P_\psi\psi',\label{matterdiff}\\
    \ham_m&:=\frac{\sqrt{|F|}}{2}\left(-\s\frac{P_\psi^2}{r^2}+r^2(\psi')^2\right)+\frac{r^2}{2\sqrt{|F|}}V(\psi).\label{matterham}
\end{align}
The total Hamiltonian of the system then reads
\begin{align}
H_T=\int\bigg(N\big(\ham+\ham_m\big)+N^x\big(\diff+\diff_m\big)\bigg)dx,
\end{align}
with $\ham$ and $\diff$ as defined in \eqref{hamSO3vacmod+} and \eqref{eq.diff}, respectively, and the matter contributions to the constraints \eqref{matterdiff} and \eqref{matterham}.
This new set of constraints satisfies the canonical hypersurface deformation algebra \eqref{eq.hdageneral} by construction.

In the GR limit, where $F=\erad/\ephi^2$ and $r=\sqrt{\erad}$, we recover the usual form,
\begin{align}\label{hammgr}
\!\!\!\!\ham_m^{\rm (GR)}=\frac{P_\psi^2}{2\sqrt{\erad}\ephi}+\frac{\erad^{3/2}(\psi')^2}{2\ephi}+\frac{\sqrt{\erad}}{2}\ephi V(\psi).\!\!
\end{align}

Certainly, this can be generalized to any Lorentz-invariant Lagrangian.
For completeness, since it could be of relevance to study gravitational collapse,
we provide here also the contributions of a minimally coupled
dust field $\phi$
to the diffeomorphism and Hamiltonian constraints of the model,
\begin{align}
    \diff_{ m}&:=P_\phi\phi',\\
    \ham_{m}&:=P_\phi\sqrt{1+|F|\phi'^2},
\end{align}
where $P_\phi$ is the conjugate momentum to $\phi$.

As can be seen in the above examples,
the resulting matter Hamiltonian is just the same as it would be in GR with minimally coupled matter, with the only difference lying on the (radial component of the) {metric}. Therefore, the minimal coupling keeps the same functional form (just replacing $q^{xx}=E^x/(E^\varphi)^2$ with $F$) while exhibiting, in principle, a different dynamical behavior.

The above result is of great relevance in the field of effective loop quantum gravity, since holonomy corrections have long been considered incompatible with the presence of
matter fields (see, e.g., Refs.{\cite{Bojowald:2015zha,Alonso-Bardaji:2020rxb}}). This was not a covariance issue as in pure vacuum,
but the matter coupling rather produced inevitable anomalous terms in the hypersurface deformation algebra. 
However, in Ref.~\cite{Alonso-Bardaji:2021tvy}, a prescription was presented in order to consistently couple
matter to holonomy corrected spherical gravity. This prescription can be understood as performing 
a canonical transformation followed by a linear combination of GR constraints.
The description of matter is therefore automatic, but the procedure gives rises
to nonminimal couplings. (The contribution of a scalar field coupled in this way
to a holonomy corrected model can be seen in Eq.~(36) of Ref.~\cite{Alonso-Bardaji:2021tvy}.)
Both procedures, the minimal coupling presented here and the nonminimal one performed in Ref.~\cite{Alonso-Bardaji:2021tvy}
\textit{covariantly} couple the scalar field to the gravitational sector, although their dynamics will, in general, differ. A specific analysis of such dynamics is left for future work.

\section{Conclusions}\label{sec.concl}

We have analyzed covariant generalizations of the Hamiltonian constraint of general relativity
under the assumption of spherical symmetry.
In this context, by covariant we mean that the dynamics generated by this generalized
Hamiltonian constraint should lead in a precise sense to define a geometry on the spacetime,
independently of gauge or coordinate choices.
More precisely, we have constructed the most general Hamiltonian constraint that obeys
the four conditions $(i)$--$(iv)$ detailed in Sec. \ref{sec.conditions}.
The first condition $(i)$ requires that the constraint should have the same derivative
structure as in general relativity, without including higher-order spatial derivatives.
Therefore, the structure of the constraint is given by \eqref{eq.hamqpvac}, where its dependence
on spatial derivatives of the variables is fixed, and there are six generic free functions,
which can depend on all the variables of the system but not on their derivatives.
The second condition $(ii)$ is just a technical assumption so that some of the
free functions are not taken to be exactly vanishing in the subsequent analysis.
In this way, we ensure that the GR Hamiltonian constraint will be contained as a
particular case of the result.

Conditions $(iii)$ and $(iv)$ are the actual nontrivial requirements that implement
the covariance of the model. On the one hand, $(iii)$ demands that the Hamiltonian
constraint \eqref{eq.hamqpvac} obeys the canonical hypersurface deformation algebra \eqref{eq.hdageneral} along with
the diffeomorphism constraint \eqref{eq.diff}, which is taken to have the same form as in GR.
In this way, this algebra, which encodes the covariance of the theory in the canonical
setting, will be closed and there will be no anomalies. In addition, from here
one can interpret, as usual, the diffeomorphism constraint as the generator of deformations
on the spatial leaf, while the Hamiltonian constraint will be the generator of deformations
in the normal direction. On the other hand, $(iv)$ requires that the structure function
that appears in the bracket between two Hamiltonian constraints should have the correct transformation
properties to be interpreted as an inverse spatial metric.
Once these two conditions are met, the spacetime metric \eqref{eq.metric}
can unambiguously be defined. For this metric, any change of coordinates in spacetime
corresponds to a gauge transformation in phase space, and therefore the geometry
is covariantly defined.

The main result of the paper is then given by Eq.~\eqref{hamSO3vacmod+}, which displays the most general
Hamiltonian constraint that obeys the four conditions specified above.
Remarkably, the implementation of conditions $(iii)$ and $(iv)$ considerably
reduces the freedom: from the six free functions of four variables in our initial ansatz \eqref{eq.hamqpvac},
one ends up with just six functions of the variable $E^x$ in \eqref{hamSO3vacmod+}. It turns out that
$E^x$ is, among the basic variables of the model, the only one that is a spacetime scalar.
Interestingly, in this constraint there are some trigonometric functions (which can also
be hyperbolic), which could be interpreted as holonomy corrections in the context
of effective-model building of loop quantum gravity. We would like to emphasize again that
in our study such trigonometric functions appear as a direct consequence of
the covariance conditions $(iii)$ and $(iv)$. Therefore, this result might be a more fundamental
motivation for holonomy corrections, and could be used as a guide to construct covariant
effective models in this context.
Let us point out that the theory of general relativity is recovered in the limit where the argument
of the trigonometric functions tends to zero, and can be understood as some kind
of limiting case between the trigonometric and hyperbolic behavior of the Hamiltonian.

The dynamics generated by the family of Hamiltonian constraints \eqref{hamSO3vacmod+} defines a geometry
in spacetime, though only for the sector ${\cal M}^2$. Due to the assumption of spherical
symmetry, the hypersurface deformation algebra does not encode the complete information
of the full four-dimensional spacetime, and, in particular, the form of the area-radius
function $r$ is missing. This adds another free function to the model. With all these
free functions, it is not possible to explicitly solve the equations of motion and obtain
the geometry. However, we have been able to conclude a number of relevant features of
the spacetimes under consideration.
In particular, we have shown, and explicitly computed,
that all the spacetimes contain a Killing vector field, in addition to the three Killing
fields of spherical symmetry.
Furthermore, unlike in GR, we observe that in general the structure function
in the bracket between two Hamiltonian constraints
is not positive definite, and thus the spacetimes can have either a Riemannian or a Lorentzian
signature. In fact, the structure function might not even be defined for all the values
of $r$, and thus its domain might be restricted
(this is what happens, for instance in the particular models studied
in Refs. \cite{Alonso-Bardaji:2021yls,Alonso-Bardaji:2022ear,Alonso-Bardaji:2023niu}).
In general, the form of the structure function will define domains of $r$,
which are either Riemannian or Lorentzian.  In the Lorentzian sector, the sign
of the Killing field defines two generic types of regions:
static nontrapped regions (similar to the exterior of a Schwarzschild black hole) and homogeneous trapped regions (similar to the interior of a Schwarzschild black hole).
Besides, there are some relevant hypersurfaces that can be either lightlike Killing horizons, minimal hypersurfaces, or boundaries between Riemannian and Lorentzian regions.

In particular, we have studied a bit in more detail a subfamily
of models, which are defined by fixing one of the free functions of the model $(\varphi=n\pi/2)$.
Under certain conditions, we have been able to show that, for this subfamily,
the signature of the spacetime is fixed and, thus, there is no possibility of a
dynamical signature change. In addition, we have analyzed the behavior
of the different fields around the vanishing point of the structure function.

A more detailed analysis of the dynamics would require fixing some of the remaining free functions.
One open question is how general is the mechanism of singularity resolution seen in the
particular models \cite{Alonso-Bardaji:2021yls,Alonso-Bardaji:2022ear,Alonso-Bardaji:2023niu},
or what are the minimum requirements on the free functions to provide such resolution.
In those cases, as already commented above, the form of the structure function reduces
the domain of $r$ and, in some cases, $r=0$ is not included, where the curvature
scalars diverge, and thus this leads to a singularity-free spacetime.
This is generic for the black-hole model analyzed in Refs.~\cite{Alonso-Bardaji:2021yls,Alonso-Bardaji:2022ear},
but when adding charge and cosmological constant, as in Ref.~\cite{Alonso-Bardaji:2023niu}, 
the resolution of the singularity requires certain restrictions on the parameters.

As the last result of the paper we point out that, since the model
is completely covariant, in order to add matter one can simply follow the usual
minimal-coupling prescription. This is a very relevant result in the
context of effective models of loop quantum gravity, since the covariant coupling
of matter to such models has been under discussion for long time.
In particular, we explicitly provide the contributions to
the Hamiltonian and diffeomorphism constraints of a scalar and
a dust matter fields. The model can then be used to study dynamical scenarios
like a gravitational collapse. However, there are other proposals
to (nonminimally) couple matter \cite{Alonso-Bardaji:2021tvy}, that should be valid at least for certain
subfamily of the models presented here. This coupling will, in general, generate
a different dynamics and a detail comparison between both could be of interest.

Finally, we would like to comment that a similar study was recently presented
in Ref. \cite{Bojowald:2023xat} and,
as detailed in Appendix~\ref{app.bd}, the Hamiltonian \eqref{hamSO3vacmod+} is equivalent to the
one obtained in that paper.

\section*{Acknowledgments}

The authors thank Martin Bojowald, Suddhasattwa Brahma, and Ra\"ul Vera for interesting discussions,
and Erick I. Duque for correspondence.
This work was supported by the Basque Government Grant IT1628-22, and by the Grant PID2021-123226NB-I00 (funded by MCIN/AEI/10.13039/501100011033 and by “ERDF A way of making Europe”). A.A.B.'s work was made possible through the support of the ID\# 62312 grant from the John Templeton Foundation, as part of the project \href{https://www.templeton.org/grant/the-quantum-information-structure-of-spacetime-qiss-second-phase}{``The Quantum Information Structure of Spacetime'' (QISS)}. The opinions expressed in this work are those of the authors and do not necessarily reflect the views of the John Templeton Foundation.

\appendix

\section{Birkhoff's theorem}

In Sec.~\ref{sec:killing} we have derived the conditions of the existence of a Killing field
in the ${\cal M}^2$, and shown that these are obeyed by the family of Hamiltonians \eqref{hamSO3vacmod+}.
Let us now rewrite these conditions in terms of the Einstein tensor.
For the spherically symmetric spacetimes under consideration,
the Einstein tensor has four nontrivial components: three in the ${\cal M}^2$
sector,
\begin{equation}\label{einsteinAB}
{}^{(4)}G_{AB}=-\frac{2}{r}\,\nabla_{\!A}v_{B}+\frac{g_{AB}}{r^2}\left(\hs-1+2\,r\, \nabla_{\!C}v^C\right),
\end{equation}
and one in the $S^2$ sector,
\begin{equation}
{}^{(4)}G_{\theta}{}^\theta={}^{(4)}G_{\phi}{}^\phi=-\frac{R}{2}+\frac{1}{r}\nabla_Av^A.
\end{equation}
From Eq.~\eqref{einsteinAB} one can write the derivative of the vector $v_A$
as follows,
\begin{equation}\label{eq.einst}
 \nabla_{\!A}v_{B}=-\frac{r}{2}\,{}^{(4)}G_{AB}+\frac{g_{AB}}{2r}\left(1-\hs+r^2\, {}^{(4)}G_{C}{}^C\right).
\end{equation}
This relation can then be used, on the one hand, to write the condition for the existence
of the Killing vector field \eqref{eq:killingcondition} as,
\begin{equation}
{}^{(4)}G_{AB}v^{A} u^{B} =0.
\end{equation}
On the other hand, the function $h(r)$
that defines the Killing vector $\xi^A=h(r) u^A$ can be rewritten as,
\begin{equation}
 \!\!h(r)=\exp\Bigg[\int dr \frac{r}{\hs^2}\bigg( {}^{(4)}G_{AB}v^{A}v^{B}-\frac{\hs}{2} {}^{(4)}G_{A}{}^{A}\bigg)  \Bigg],
\end{equation}
as one can check by substituting \eqref{eq.einst} in \eqref{eq:hsolution}.
Now, from the last two expressions it is explicit that in vacuum
general relativity, $G_{AB}=0$, the Killing vector exists and, more precisely,
it is given by $\xi_A=u_A$.

\section{Equivalence with the model by Bojowald and Duque}\label{app.bd}

Very shortly before the submission of this paper,
the preprint \cite{Bojowald:2023xat} appeared online with a similar study. This research,
done in parallel to ours
(which is mainly based in Ref. \cite{AlonsoBardaji:2023bww}),
considers also derivatives of the extrinsic-curvature variables that are not implemented
in our ansatz \eqref{eq.hamqpvac}. However, a canonical transformation is then
performed in order to absorb those derivatives, and, their final result,
given by Eq. (147) of Ref. \cite{Bojowald:2023xat},
is indeed equivalent to the constraint \eqref{hamSO3vacmod+} derived in the present paper.
More precisely,
considering the canonical transformation from the variables
$(E^x,K_x,E^\varphi, K_\varphi)$ in Eq. \eqref{hamSO3vacmod+}
to $(\widetilde E^x,\widetilde K_x,\widetilde E^\varphi, \widetilde K_\varphi)$
as given by,
\begin{align*}
\bar{\lambda} \widetilde{K}_\varphi&=\omega(E^x) K_\varphi+\varphi(E^x),\\
\frac{\widetilde{E}^\varphi}{\bar{\lambda}}&=\frac{E^\varphi}{\omega(E^x)},\\
\widetilde{E}^x &=E^x,\\
\widetilde{K}_x&=K_x+\frac{E^\varphi}{\omega(E^x)}\left[\frac{\partial\omega(E^x)}{\partial E^x} K_\varphi+ \frac{\partial\varphi(E^x)}{\partial E^x}\right],
\end{align*}
with $\bar{\lambda}$ being a constant, and the following redefinition of our six free functions
$({\mathfrak g}, \omega, \varphi, A, V, W)$,
\begin{align*}
 \frac{\lambda_0}{\bar{\lambda}}&=\frac{{\mathfrak{g}}}{\sqrt{E^x}\omega},\\
 c_f&= A \cos(2\varphi),\\
 {\bar\lambda} q &= -\frac{A}{2}\sin(2\varphi),\\
 {\bar\lambda^2}c_{f0}&=\frac{2 A}{W}\frac{\partial W}{\partial E^x}\sin^2(\varphi),\\
 {\bar\lambda^2}\alpha_0&=\frac{2 E^x}{\omega}\left( 
 A \sin(2\varphi) \omega \frac{\partial\varphi}{\partial E^x} + \sin^2(\varphi)\omega^3 \frac{\partial}{\partial E^x}\left(\frac{A}{\omega^2}\right)-\omega^3 V \right),\\
 \alpha_2 &= 2 E^x \left(\frac{\partial\log W}{\partial E^x}-2\frac{\partial\log\omega}{\partial E^x} \right),
\end{align*}
the constraint \eqref{hamSO3vacmod+} takes the form given by Eq. (147) of Ref. \cite{Bojowald:2023xat}.

\providecommand{\noopsort}[1]{}\providecommand{\singleletter}[1]{#1}%

\end{document}